\documentclass[aps,pre,twocolumn,showpacs,superscriptaddress,floatfix]{revtex4}
\usepackage{graphicx}
\usepackage{bm}
\bibstyle{apsrev.bib}

\begin{document}
\title{Magnetophoresis in the Rubinstein -- Duke model}
\author{A. Drzewi\'nski} 
\affiliation{Institute of Low Temperature and Structure Research,
Polish Academy of Sciences, P.O. Box 1410, Wroc\l aw 2, Poland}
\author{E. Carlon}
\affiliation{Theoretische Physik, Universit\"at des Saarlandes,
D-66041 Saarbr\"ucken, Germany}
\affiliation{Interdisciplinary Research Institute c/o IEMN,
Cit\'e Scientifique BP 69, F-59652 Villeneuve d'Ascq, France}
\author{J.M.J. van Leeuwen}
\affiliation{Instituut-Lorentz, Leiden University, P.O. Box 9506,
2300 RA Leiden, The Netherlands}

\date{\today}

\begin{abstract}

We consider the magnetophoresis problem within the Rubinstein -- Duke
model, i.e. a reptating polymer pulled by a constant field applied
to a single repton at the edge of a chain.  Extensive density matrix
renormalization calculations are presented of the drift velocity and
the profile of the chain for various strengths of the driving field and
chain lengths.  We show that the velocities and the average densities
of stored length are well described by simple interpolating crossover
formulae, derived under the assumption that the difference between
the drift and curvilinear velocities vanishes for sufficiently long
chains. The profiles, which describe the average shape of the reptating
chain, also show interesting features as some non-monotonic behavior of
the links densities for sufficiently strong pulling fields.  We develop
a description in which a distinction is made between links entering at the
pulled head and at the unpulled tail. At weak fields the separation between
the head zone and the tail zone meanders through the whole chain, while the
probability of finding it close to the edges drops off. At strong fields
the tail zone is confined to a small region close to the unpulled edge
of the polymer.

\end{abstract}


\maketitle

\section{Introduction}

The magnetophoresis problems is a member of the class of reptation problems 
in which a long polymer is driven through a gel. The reptative motion of the 
polymer can be succesfully modeled by a lattice version using sections of 
the polymer (the reptons) as the mobile units, which hop stochastically 
from cell to cell.
The driving field is incorporated as a 
bias in the hopping rates, favoring the motion in the field direction. A 
simple and adequate model is the Rubistein - Duke model \cite{RD}, which 
represents the polymer as a chain of $N$ independently moving reptons, 
with the restriction that the integrity of the chain is preserved.
The reptons trace out a connected string of cells in space, each cell 
containing at least one repton. The cells, which can be multiply occupied, 
carry the extra reptons as units of stored length. In order to preserve 
the integrity of the chain, only those reptons which are located in cells 
with stored length can hop.

Usually one considers the case of polyelectrolytes in which the reptons 
are uniformly charged. Thus the driving field pulls equally on all reptons 
and the bias is the same for the hopping rates all along the chain.
A practical situation where this occurs is in DNA electrophoresis 
\cite{viovy}. The DNA molecules, being acid, get charged in solution 
and when they are placed in a gel subject to an external electric field
they perform a biased reptative motion along the field direction. 
Electrophoresis is a technique of great importance in molecular biology 
and sequence analysis, as it allows to separate DNA strands according 
to their length \cite{viovy}.

\begin{figure}[t]
\includegraphics[width=8.2cm]{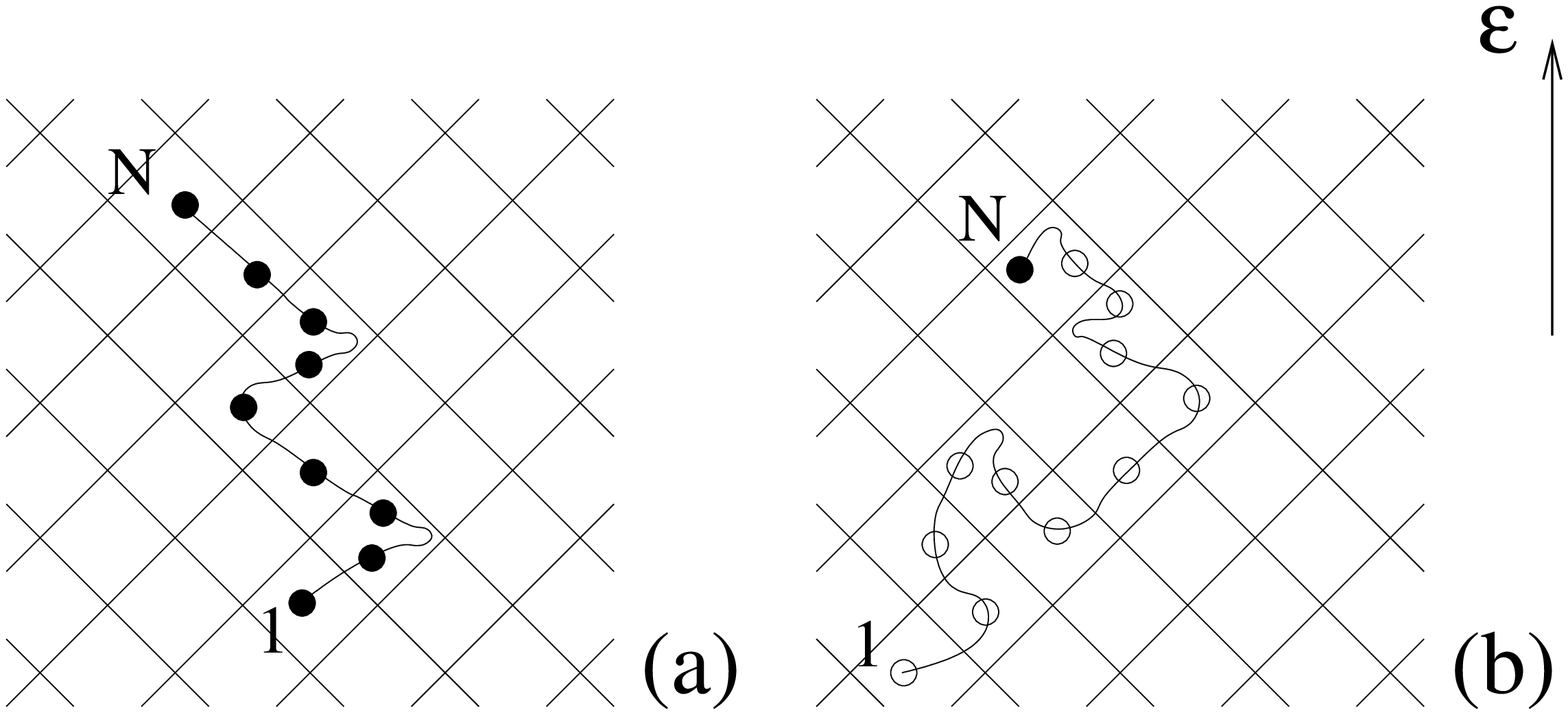}
\caption{Examples of configurations of reptating polymers in the 
Rubistein -- Duke model in the case of (a) Electrophoresis and (b)
Magnetophoresis. Black reptons perform a biased motion along the
direction of the applied field while white reptons are unbiased.
The configuration for a chain with $N$ reptons is given by a set of 
$N-1$ integers $(y_1, y_2 \ldots y_{N-1})$
measuring the distance of two neighboring reptons along the field
direction (thus $y_i = 0, \pm 1$). For the two examples shown the
coordinates are $(1,0,1,1,1,0,1,1)$ for (a) and 
$(1,1,1,0,-1,1,1,1,0,1,0)$ for (b).
}
\label{RD}
\end{figure}

In nature the  charge distribution is of course 
not always uniform. The extreme alternative is the case where all reptons 
are neutral except one end repton, which  is charged. A possible realization 
of such a situation is a magnetic bead, attached at one end of the polymer, 
which is driven by a magnetic field. Therefore this case can be referred 
to as the magnetophoresis (MP) problem \cite{bark96}. 
It is the subject of this paper. As we will frequently compare our findings 
with the more common case of the uniformly charged polymer, we will, for 
briefness, refer to the latter as the electrophoresis (EP) problem and to 
the present one as the MP problem, although the distinction between the two 
is not electromagnetic, but only in the forces exerted on the reptons
(see Fig. \ref{RD}).

%%%%%%%% FIG 1  %%%%%%%%%%%%%%%%%%%%%%%%%%%%%%%%%%%%%%%%%%%%%%%%%%%%%%%%%
%%%%%%%%%%%%%%%%%%%%%%%%%%%%%%%%%%%%%%%%%%%%%%%%%%%%%%%%%%%%%%%%%%%%%%%%%

The MP problem, within the framework of the Rubinstein--Duke model, has 
so far been studied by means of Monte Carlo simulations \cite{bark96}, 
and the calculations were mostly restricted to the drift velocity as 
function of the applied field and chain length.
In this paper we analyze the MP problem by means of density-matrix 
renormalization-group (DMRG) techniques, which allow us to perform
a detailed analysis of both global quantities as drift and curvilinear
velocities and diffusion coefficient, but also on local average shapes
of the polymer.

The dynamics of the reptating chain is governed by the Master Equation 
which we put in the form
\begin{equation} 
{\partial P ({\bf y}, t) \over \partial t} = 
\sum_{\bf y'} H({\bf y}, {\bf y'}) P ({\bf y'}, t)
\label{aa1}
\end{equation} 
Here ${\bf y}$ stands for the set of links $y_1, \cdots ,y_{N-1}$, where
$y_j$ measures the distance between the repton $j$ and $j+1$, along
the direction of the applied field.  In our lattice representation the
$y_j$ can take the values $\pm 1$ and $0$. The value $y_j=0$ corresponds
to the case that the reptons $j$ and $j+1$ occupy the same cell. Thus
each zero is a unit of stored length.  The non--zero values represent
the cases where $j+1$ occupies a cell ``higher'' (1) or ``lower'' (-1)
than $j$. Higher and lower refer to a position in the direction of the
field. ${\bf y}$ represents a complete configuration of the chain (see
Fig. \ref{RD}).  $P ({\bf y}, t)$ is the probability distribution of
the configuration at time $t$ and the matrix $H({\bf y, y'})$ contains
the gain and loss transitions from ${\bf y'}$ to ${\bf y}$. The bias in
the hopping rate is contained in the matrix elements of $H$. Generally
we have for the bias factor
\begin{equation} 
B_j = \exp (aq_j E/k_B T)
\label{aa2}
\end{equation} 
with $q_j$ the charge of repton $j$, $E$ the driving field, $a$ the 
distance between adjacent cells (measured along the field direction) 
and $k_B T$ the standard combination of Boltzmann's constant and the 
absolute temperature. In the MP problem all $q_j = 0$ except for $j=N$. 
We put
\begin{equation} 
B_{N} \equiv B = \exp (\varepsilon/2)
\label{aa3}
\end{equation} 
and use $\varepsilon$ as the parameter for the driving field. The other 
parameter of the model is the number of reptons N. 
  
We have chosen this ``hamiltonian form'' of the Master Equation in
order to stress the formal correspondence with a quantum mechanical model
governed by a hamiltonian matrix.  The DMRG method exploits this analogy
and indeed its success in one-dimensional quantum problems carries over to
reptation problems \cite{paper1,paper2,paes02}.  We are interested in the
stationary state of the probability distribution.  In the quantum language
this corresponds to finding the right eigenvector of $H$ belonging to
the eigenvalue zero. The adaptation of the DMRG method to the MP problem
is straightforward and the data presented in this paper are obtained
by the DMRG method.  An important difference with the Master Equation
is that in quantum mechanical problems the hamiltonian is hermitian,
whereas in the reptation problem the matrix $H$ is non--hermitian, due
to the influence of the driving field. This restricts the applicability
of the DMRG--method to moderately long chains and/or small driving fields.

The physics of the EP and MP problems is qualitatively different. As
illustration consider the weak field case ($\varepsilon$ small), where
the Nernst--Einstein relation $v = F D$ relates the drift velocity
$v$ to the total applied force $F$ and the diffusion coefficient $D$.
The force $F$, equals $N \varepsilon$ in the EP case, since one pulls at
each repton. It is well known that $v$ scales as $v \sim \varepsilon /N$,
yielding for the diffusion the non--trivial result $D \sim N^{-2}$. In
the MP problem $D$ will be essentially the same, as hopping is limited
by the availability of stored length. In both cases the motion can be
considered as diffusion of stored length. Since $F=\varepsilon$ in the
MP problem, we expect the drift velocity to scale as $v \sim \varepsilon
N^{-2}$, a feature which is born out by our calculations.

In the Sec. \ref{sec:moments} we discuss some moment equations derived
from the Master Equation which are more helpful then in the EP problem
in analyzing the drift-- and curvilinear velocity.
They are expressed in terms of the
probabilities $n^k_j$ that the link $j$ has the value $y_j=k$. The sum
of the $n^k_j$ adds up to 1
\begin{equation} 
n^0_j + n^+_j + n^-_j = 1
\label{aa4}
\end{equation} 
So it suffices to consider the two quantities $n^0_j$ and $m_j$ defined by
\begin{equation} 
n^0_j = \langle \, 1 - y^2_j \, \rangle \quad \quad \quad 
m_j = \langle \, y_j \, \rangle = n^+_j - n^-_j
\label{a5}
\end{equation} 
$n^0_j$ can be called the local density of stored length. $m_j$ is a
measure for  the local orientation and will be  referred to as the 
profile of the chain.

In the Sections \ref{sec:global} and \ref{sec:profiles} we present
our data for the velocities and the profiles.  The analysis is most
transparent in the strong field limit where we can make an ansatz which
almost perfectly represents the data. In section \ref{sec:profiles} we 
discuss the behavior of the profile for weak and strong pulling fields.

\section{Moments of the Master Equation}
\label{sec:moments}

The DMRG method deals with the whole probability distribution $P({\bf
y})$. In the MP problem it is fruitful to consider moments of the Master
Equation. One set of moments is obtained by multiplying Eq. (\ref{aa1})
with $y_j$ and then summing over all ${\bf y}$. This leads to $N-1$
relations which can be seen as an expression of the fact that the drift
velocity $v$ across all the $N-1$ links of the chain must be the same
on the average. An even more useful set of relations is obtained by
multiplying the Master Equation by $y^2_j$ and summing over all ${\bf
y}$. The resulting $N-1$ relations are an expression of the fact that the
curvilinear velocity $J$ is the same across all links. These relations
obtain the form
\begin{equation} 
J = n^0_{j-1} - n^0_j \quad \quad \quad 1 < j < N 
\label{bb1}
\end{equation} 
which can be viewed as the familiar law that the current $J$ equals
minus the gradient of the density of the stored length. In addition to
Eq. (\ref{bb1}) one has two relations \cite{footnote} concerning the
traffic in and out both ends of the chain. They read
\begin{eqnarray} 
J &=& 1 - 3 n^0_1, \nonumber\\
J &=& (n^0_{N-1} - n^-_{N-1}) B + (n^0_{N-1} - n^+_{N-1}) B^{-1}
\label{bb2}
\end{eqnarray} 

The expression for the drift velocity involves correlations between 
neighboring links
\begin{equation} 
v = \langle \, (1- y^2_{j-1}) y_j - (1 - y^2_j ) y_{j-1} \, \rangle
\label{bb3}
\end{equation} 
and it is therefore not as informative as Eq. (\ref{bb1}). 
As one sees Eq. (\ref{bb2}) involves only averages
over the first (last) link. This holds also for the expressions for the
drift velocity in terms of the averages of the first and last link.
\begin{eqnarray} 
v &=& m_1, \nonumber \\ %\quad \quad \quad 
v &=& (n^0_{N-1} + n^-_{N-1}) B - (n^0_{N-1} + n^+_{N-1}) B^{-1}
\label{bb4}
\end{eqnarray}   
These equations have been derived by Barkema and Sch\"utz \cite{bark96}
using balance arguments.

Equation (\ref{bb1}) is a powerful relation since it allows to express 
the density of stored length in terms of the curvilinear velocity
\begin{equation} 
n^0_j = n^0_1 - (j-1) J
\label{bb5}
\end{equation}
showing that the density profile is linear in the position of the cell. 
In particular (\ref{bb5}) implies a relation between the densities of 
the first and last cell
\begin{equation} 
n^0_{N-1} = n^0_1 - (N-2) J
\label{bb6}
\end{equation} 
We have used the linearity of the density $n^0_j$ as a check of 
the numerical calculations.

Counting the number of unknowns ($v$, $J$, $n^0_1$, $n^0_{N-1}$,
$m_1$, $m_{N-1}$) and the number of equations  (\ref{bb2}), (\ref{bb4})
and (\ref{bb6}) we see that we have one more unknown than equations.
This situation is similar to the EP problem. There the expression for the
curvilinear velocity does not obtain the simple form (\ref{bb1}), due to
the bias on the internal reptons. So one misses relation (\ref{bb6}).
On the other hand $J=0$ for EP, due to the symmetry of the polymer on
exchanging head and tail. Thus in both cases the moment equations are
not sufficient to determine the velocities. Higher moments do not lead
to additional information since again higher order correlations appear.

\section{Global quantities}
\label{sec:global}

We discuss first the behavior of global quantities as the drift and curvilinear
velocities and the diffusion constant.

\subsection{The Weak Field Limit}

\begin{figure}[t]
\centering
\includegraphics[width=8.5cm]{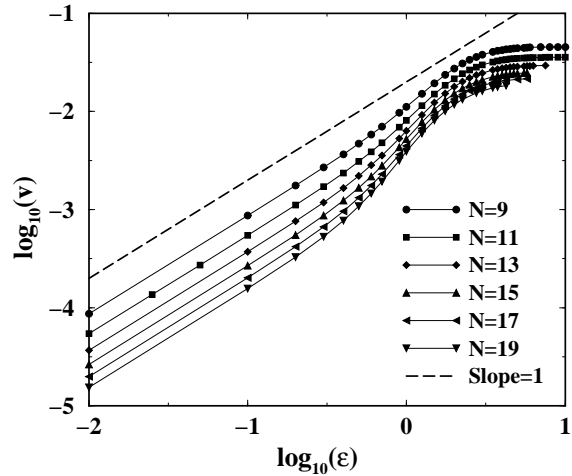}
\caption{ Plot of $\log v$ vs. $\log \varepsilon$ for various $N$.}
\label{fig02}
\end{figure}

In the weak field limit the polymer assumes mostly a random configuration 
and all the densities $n^k_j$ are close to 1/3. The overal behavior of the
drift velocity $v$ as function of $\varepsilon$ and $N$ is given in 
Fig. \ref{fig02}. Note that the drift velocity becomes proportional to 
$\varepsilon$ for small $\varepsilon$, as expected on the basis of the 
Nernst -- Einstein relation discussed above. For stronger fields, the
velocity saturates to a finite values, as discussed in the next paragraph.
A similar dependence of $v$ on $\varepsilon$ and $N$ was observed in 
the Monte Carlo study of Ref. \cite{bark96}.
The limiting behavior for small $\varepsilon$ and large $N$ is thus:
\begin{equation} \label{c1}
v(N) \sim \varepsilon D(N) \sim \frac{\varepsilon}{N^2},
\end{equation}
with $D(N)$ the zero field diffusion coefficient.
The scaling behavior of $D(N)$ as function of the length $N$ has been
studied quite intensively \cite{wido91,bark97,bark98,fris00}. Reptation
theory predicts that $D(N) \sim 1/N^2$, while conflicting results appeared
on experimental measurements, for which both $1/N^2$ and, more recently,
$1/N^{2.3}$ \cite{lodge} have been reported.

%%%%%%%% FIG 2  %%%%%%%%%%%%%%%%%%%%%%%%%%%%%%%%%%%%%%%%%%%%%%%%%%%%%%%%%
%%%%%%%% FIG 2  %%%%%%%%%%%%%%%%%%%%%%%%%%%%%%%%%%%%%%%%%%%%%%%%%%%%%%%%%

A detailed study of the scaling of $D(N)$, within the Rubinstein--Duke
model by means of DMRG method was recently performed \cite{paper1,paper2},
for various end-point stretching rates. In that case the diffusion 
coefficient was calculated from the limiting value of the drift velocity
for $\varepsilon \to 0$, with the field acting on all reptons (the EP problem).
Here we repeat the same analysis only for a single case (using a stretching 
rate $d=1$ following the definition of $d$ of Refs. \cite{paper1,paper2}).
The advantage of calculating $D(N)$ with a small field acting only on an
end repton is that the DMRG procedure is much more stable in this case and
one can compute longer chains. This is due to the fact that in the MP
problem non-hermiticity is restricted only to the repton where the field
is applied. As mentioned in the introduction non-hermiticity hampers the
efficiency of the DMRG method. 

In order to calculate the diffusion coefficient from the Nernst-Einstein
relation $D = \lim_{\varepsilon \to 0} v/\varepsilon$ in practice, 
we used a small field ($\varepsilon = 10^{-3}$) and checked 
explicitely that results do not change for smaller fields. The scaling 
behavior of the diffusion coefficient is expected to be:
\begin{equation} \label{c3}
D(N) N^2 =  A + \frac{A'}{\sqrt{N}} + \ldots
\end{equation}
with $A$ and $A'$ some constants. The form of the subleading correction to $D(N)$ 
has been debated for a while \cite{bark,prah} and recent DMRG results 
suggest that it is of the type $1/\sqrt{N}$ \cite{paper1}, supporting 
Eq. (\ref{c3}). The coefficient was determined exactly \cite{vLK,prah, Al-L}:
$A = 1/3$.

\begin{figure}[t]
\centering
\includegraphics[width=8.5cm]{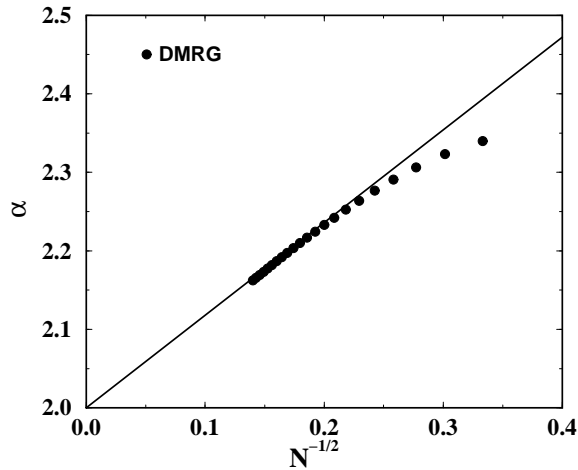}
\caption{Plot of the effective exponent $\alpha(N)$ calculated for 
  $\varepsilon=10^{-3}$ up to $N=51$ from the decay of the 
  diffusion coefficient $D(N)$ and plotted as a function of 
  $1/\sqrt{N}$. The fact that $\alpha(N)$ approaches linearly
  the limiting value $2$ supports the scaling form for the
  diffusion coefficient given in Eq.(13).}
\label{diff}
\end{figure}

To analyze the scaling behavior of $D(N)$ it is most convenient to use the
logarithmic derivative of the DMRG data:
\begin{equation} \label{c2}
\alpha (N) =  - \frac{ \ln \left[ D (N) \right]
- \ln \left[ D (N+2) \right]}{\ln  N-\ln (N+2)} ,
\end{equation}
which is shown for $N =9$, $11$, \ldots, $51$ in Fig. \ref{diff}. 
Plugging 
Eq. (\ref{c3}) in Eq. (\ref{c2}) one finds for the effective exponent
$\alpha (N) = 2 + A'/(2 A \sqrt{N})$, a behavior which is accurately
reproduced by our numerical data of Fig. \ref{diff}.  The present
results corroborate previous claims \cite{paper1} about the scaling form
of $D(N)$.

%%%%%%%% FIG 3  %%%%%%%%%%%%%%%%%%%%%%%%%%%%%%%%%%%%%%%%%%%%%%%%%%%%%%%%%
%%%%%%%% FIG 3  %%%%%%%%%%%%%%%%%%%%%%%%%%%%%%%%%%%%%%%%%%%%%%%%%%%%%%%%%

\subsection{The Strong Field Limit}

%%%%%%%% FIG 4  %%%%%%%%%%%%%%%%%%%%%%%%%%%%%%%%%%%%%%%%%%%%%%%%%%%%%%%%%
\begin{figure}[t]
\centering
\includegraphics[width=8.5cm]{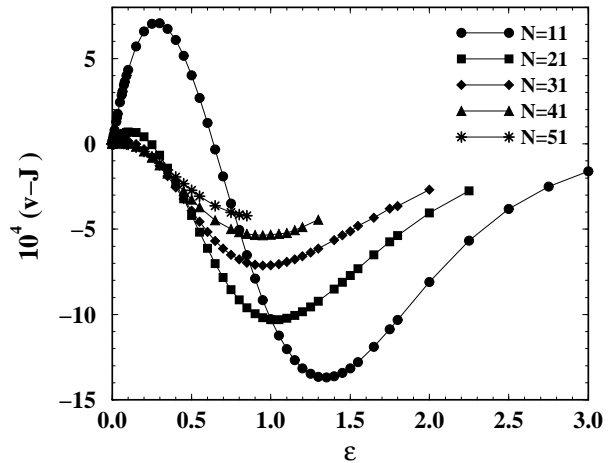}
\caption{Difference between the drift $v$ and curvilinear $J$ velocities
as function of the applied field and for various chain lengths.}
\label{diffvJN}
\end{figure}
%%%%%%%% FIG 4  %%%%%%%%%%%%%%%%%%%%%%%%%%%%%%%%%%%%%%%%%%%%%%%%%%%%%%%%%

%%%%%%%% FIG 5  %%%%%%%%%%%%%%%%%%%%%%%%%%%%%%%%%%%%%%%%%%%%%%%%%%%%%%%%%
\begin{figure}[b]
\centering
\includegraphics[width=8.5cm]{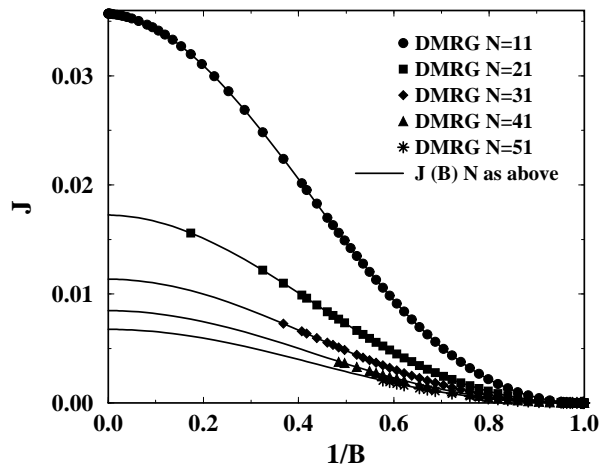}
\caption{Comparison between the DMRG data (symbols) and the crossover
formula (solid lines) of Eq. (\ref{d5}) for the curvilinear velocity $J$.}
\label{JNcrov}
\end{figure}
%%%%%%%% FIG 5  %%%%%%%%%%%%%%%%%%%%%%%%%%%%%%%%%%%%%%%%%%%%%%%%%%%%%%%%%

\begin{figure}[t]
\centering
\includegraphics[width=8.5cm]{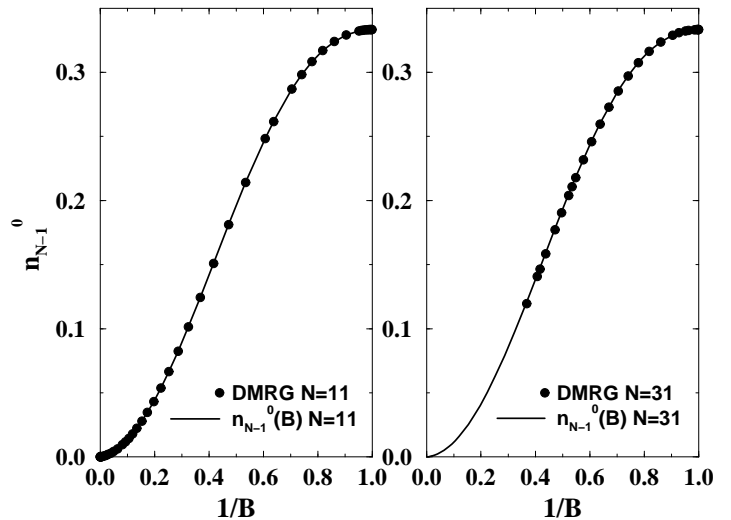}
\caption{Comparison between the DMRG data (symbols) and the crossover
formula (solid lines) of Eq. (\ref{d5b}) for $n^0_{N-1}$.}
\label{n0Ncrov}
\end{figure}

In the strong field limit the polymer assumes an oriented configuration, 
with the '+' links dominating at the pulled end. At the other end we still 
have a substantial amount of links '0', since the polymer can only move by 
the diffusion of stored length from the tail to the head. Eliminating $n^0_1$ 
from Eq. (\ref{bb5}) with the use of Eq. (\ref{bb2}) we get
\begin{equation} \label{d1}
n^0_{N-1} = {1 \over 3} (1 - K J), 
\end{equation} 
where $K = 3N - 5$.
In order that $n^0_{N-1}$ stays finite for $N \rightarrow \infty$, the 
curvilinear velocity must vanish as
\begin{equation} \label{d2}
J \sim K^{-1} \quad \quad \quad N \rightarrow \infty
\end{equation} 
As one sees from (\ref{d1}) this limiting value is not sufficient to 
determine the limiting value of density $n^0_{N-1}$, which is sensitive 
to the corrections to (\ref{d2}). For the strong field limit it is useful 
to relate the drift velocity to the curvilinear velocity. With (\ref{bb2}) 
and (\ref{bb3}) we get
\begin{equation} \label{d3}
v=J+2(n_{N-1}^{-} B - n_{N-1}^{0} B^{-1}),
\end{equation}
Now, if the polymer is fully stretched, $v$ and $J$ become the same. 
In Fig. \ref{diffvJN} we have plotted the difference $v-J$ as calculated by
DMRG for various fields and chain lengths. We note that it is small for all 
values of $\varepsilon$ and $N$, in particular for strong fields and that 
this tendency is enforced for long polymers. That it also is small in the 
small field limit is a consequence of the fact that both quantities vanish 
in that limit.
In order for the difference to vanish  we must have
\begin{equation} \label{d4}
n^0_{N-1} \simeq B^2 n^-_{N-1}
\end{equation} 
Now we may use this relation as the 6th relation, which enables us to make all the desired 
quantities explicit functions of $\varepsilon$ and $N$. We find for instance
\begin{eqnarray} \label{d5}
J(B)=v(B)=\frac{B^4-2 B^2 +1}{K(B^4+B^2+1)+3 B^3},\\
n_{N-1}^{0}(B)=\frac{1+B/K}{B^2+1 + 3B/K+1/B^2}.
\label{d5b}
\end{eqnarray}
This explicit field dependence is compared to the data in Fig. \ref{JNcrov} 
and Fig. \ref{n0Ncrov}. 
The agreement is excellent in both cases. Note also that Eq. (\ref{d5}) is consistent 
with Eq. (\ref{d2}) and that it provides the proportionality coefficient. 

However, the crossover formulae (\ref{d5})-(\ref{d5b}) do not describe the
subtile dependencies in the limit of small fields $B =
\exp(\varepsilon/2) \to 1$.  In this limit the drift 
velocity vanishes as $v \sim \varepsilon$, while one observes from (\ref{d5}),
that the curvilinear velocity vanishes as $J \sim \varepsilon^2$. 
For this reason, in the limit $\varepsilon \to 0$, the
crossover formula (\ref{d5}) predicts $v \sim \varepsilon^2/(3 N - 5)$,
in disagreement with the correct scaling behavior of Eq. (\ref{c1}).
The strong field limit does not suffer from this problem. We note, for
instance, that in the limit $N \to \infty$ the saturation value of the
velocity for $B \to \infty$ is in agreement with the exact expression
given in Ref. \cite{bark96}: $v = 1/(3 N - 5)$.  

%%%%%%%% FIG 6  %%%%%%%%%%%%%%%%%%%%%%%%%%%%%%%%%%%%%%%%%%%%%%%%%%%%%%%%%
%%%%%%%% FIG 6  %%%%%%%%%%%%%%%%%%%%%%%%%%%%%%%%%%%%%%%%%%%%%%%%%%%%%%%%%

\section{Profiles}
\label{sec:profiles}

\begin{figure}[t]
\centering
\includegraphics[width=8.5cm]{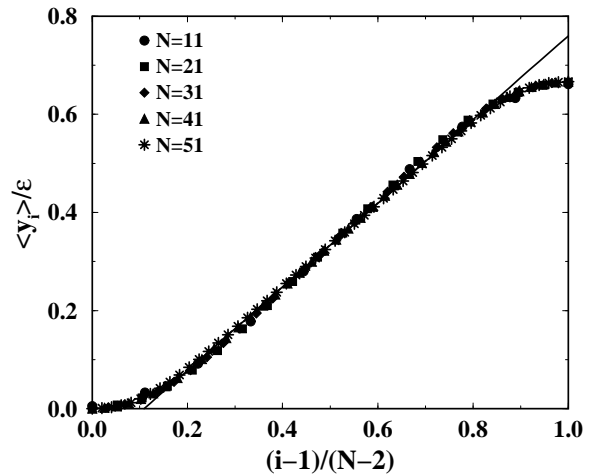}
\caption{Average profiles $\langle y_i \rangle / \varepsilon$ for various
lengths and for $\varepsilon = 10^{-3}$.}
\label{profE0-001}
\end{figure}

\begin{figure}[b]
\centering
\includegraphics[width=8.5cm]{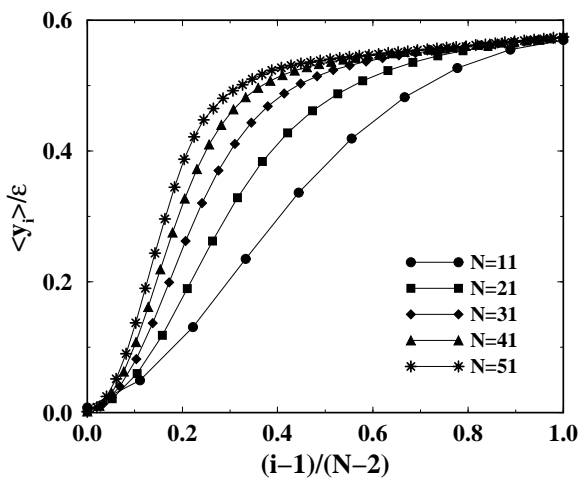}
\caption{As in Fig. \protect\ref{profE0-001} for $\varepsilon = 1$.}
\label{profE1}
\end{figure}

Next we discuss some profiles, {\bf i.e.} the local orientation
$m_i \equiv \langle y_i \rangle$ as function of the segment position
along the chain.  We consider $N-1$ segments, thus $N$-reptons with the
charged one at head position $N$.  Fig. \ref{profE0-001} shows a plot
of $m_i/\epsilon $ as function of the scaled variable $(i-1)/(N-2)$ for
chains of various lengths and at fixed field $\varepsilon = 0.001$. This
profile corresponds to the linear regime where the drift velocity scales
as $v \sim \varepsilon$. The notable feature is a symmetry 
between head and tail with
respect to the center of the chain, although the magnetophoresis problem
is clearly asymmetric. This symmetry can be shown \cite{next} to be strict in the weak field limit. 
It disappears for stronger values of
the field as Fig. \ref{profE1} shows, where profiles are plotted for
$\epsilon = 1$ and various lengths N. 

%%%%%%%% FIG 7  %%%%%%%%%%%%%%%%%%%%%%%%%%%%%%%%%%%%%%%%%%%%%%%%%%%%%%%%%
%%%%%%%% FIG 7  %%%%%%%%%%%%%%%%%%%%%%%%%%%%%%%%%%%%%%%%%%%%%%%%%%%%%%%%%

%%%%%%%% FIG 8  %%%%%%%%%%%%%%%%%%%%%%%%%%%%%%%%%%%%%%%%%%%%%%%%%%%%%%%%%
%%%%%%%% FIG 8  %%%%%%%%%%%%%%%%%%%%%%%%%%%%%%%%%%%%%%%%%%%%%%%%%%%%%%%%%

\begin{figure}[t]
\centering
\includegraphics[width=8.5cm]{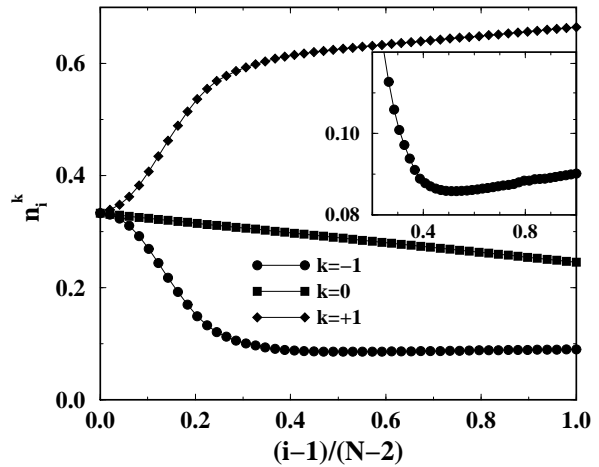}
\caption{Plot of the average densities $\langle n^+_i \rangle$,
$\langle n^0_i \rangle$ and $\langle n^-_i \rangle$ for $\varepsilon
= 1$ and $N=51$. Inset: Blow up of the density $n^-_i$, showing
a non-monotonic behavior with a linear increase as function of $i$
by approaching the pulled edge.} 
\label{nk_inter} 
\end{figure}

In order to analyze the data further we also plot the individual
probabilities $n^k_i$ for having a $+, 0$ or $-$ at the site $i$
of the chain. For small values of $\epsilon$ (not shown here) the curves 
are all near
$1/3$, with a slight excess of $+$ links at the head and a depletion
of $-$ links.  The densities of $+$ and $-$ links are monotonically 
increasing and decreasing functions of the position $i$.
For intermediate fields $\epsilon = 1$, the densities are more interesting 
and in Fig. \ref{nk_inter} we plot the values of $n^0_i$, $n^+_i$ and
$n^-_i$ for $N=51$. The linear behavior for $n^0_i$ 
is consistent with Eq. (\ref{bb5}).
The curve for $n^+_i$ is monotonically increasing, but that for
$n^-_i$ is {\it not} monotonically decreasing.

%%%%%%%% FIG 9  %%%%%%%%%%%%%%%%%%%%%%%%%%%%%%%%%%%%%%%%%%%%%%%%%%%%%%%%
%%%%%%%% FIG 9 %%%%%%%%%%%%%%%%%%%%%%%%%%%%%%%%%%%%%%%%%%%%%%%%%%%%%%%%

The qualitative behavior of the orientation profile can be understood
by considering the ``origin'' of the non-zero links ($y_i = \pm 1$)
as has been introduced by Barkema and Newman \cite{bark97}. In the MP
problem more links are created at the pulled head than at the tail.
They stream gradually down to the tail. We can keep track for every link
$y_i = \pm 1$, whether it is formed at the head or at the tail. After
sufficient time the chain is divided into two zones: a head zone and a
tail zone. They are separated by a small intermediate region with zeroes
(we do not follow the origin of the zeroes). The zones remain separated
because the $y_i = \pm 1$, created at the head cannot cross the  $y_i =
\pm 1$ created at the tail. The division between the two zones fluctuates
in time and occasionally the tail zone disappears, while very rarely
(particularly at large fields) the head zone vanishes.  The larger
the force on the head, the larger the asymmetry between the head and
tail zones.  We supplement these speculations by making an assumption
on the ratios

\begin{equation} 
\label{a3} 
r_i (j)  = p^+_i (j)  / p^-_i (j) 
\end{equation} 
where $ p^\pm_i (j)$ is the probability of finding a $\pm$ at $i$ when
the division is at $j$.  We put 
\begin{eqnarray} 
r_i (j) = r_h & \quad {\rm for} \quad j < i  \\
r_i (j) = 1 & \quad {\rm for} \quad j \geq i
\label{a4} 
\end{eqnarray}
Since in the tail zone there is no distinction between $+$ and $-$ we
have set the ratio equal to 1. The idea, underlying this assumption,
is that the $+$ and $-$ links are interlocked.  So while moving in their
zone their ratio can not change.

At position $i$ the average number of nonzero links ($+$ and $-$)
is equal to $1-n^0_i$. We introduce $f_i$ as the fraction of such nonzero
links which are in the head zone. 
One can express the densities $n^\pm_i$ in terms of $f_i$ as

\begin{equation} 
\left\{ \begin{array}{rcl}
n^+_i  & = & \displaystyle \left(f_i \, {r_h \over r_h +1}  +
(1-f_i) \,{1 \over 2} \right) [1 - n^0_i ]  \\*[4mm]
n^-_i  & = & \displaystyle \left( f_i \,{1 \over r_h +1} \, +
(1-f_i) \, {1 \over 2} \right) [1 - n^0_i ] 
\end{array} \right.
\label{b4}
\end{equation}
In both equations the terms proportional to $f_i$ are the contributions that
$i$ is in the head zone while the terms proportional to $1-f_i$ are the
contributions from the case in which $i$ is in the tail zone.
We see that in the magnetophoresis
problem the situation simplifies, since we do not have to worry about
the tail zone. It drops out when we consider the profile
\begin{equation}
m_i = n^+_i  - n^-_i  = {r_h -1 \over r_h +1} f_i \, [1- n^0_i]
\label{b5} 
\end{equation} 
Thus the profile $m_i$ is, apart from the known factor $1- n^0_i$,
directly related to the fraction $f_i$. The latter has a simpler
interpretation. It starts out at $i=1$ with a value nearly zero, since the
head zone will only seldomly extend over the whole chain. It ends at $i=N-1$
at a value very close to 1, since the tail zone will hardly ever extend over the whole
zone. We can use this fact to tie the ratio $r_h$ to the end point values,
discussed earlier, by considering (\ref{b5}) for $i=N-1$
\begin{equation} 
\label{b9}
\langle \, y_{N-1} \, \rangle = {r_h - 1 \over r_h +1} [1 - n^0_{N-1}]
\end{equation}
and solving for $r_h$. It leads to
\begin{equation} 
\label{b10}
\langle \, y_i \, \rangle  = 
\langle \, y_{N-1} \, \rangle f_j \,{1 - n^0_i \over 1 - n^0_{N-1}}
\end{equation}
This form contains only $f_i$ as unknown. We can
draw some conclusions from Eqs. (\ref{b5}) and (\ref{b10}) for weak fields, as well 
as for long chains at stronger fields.

\subsection{Weak Fields}

For $\epsilon \rightarrow 0$ we may put
\begin{equation} \label{b6}
r_h = 1 + a_h \epsilon
\end{equation}
The function $1-n^0_i$ will approach the limit $2/3$, so (\ref{b5}) becomes
\begin{equation} 
\langle \,y_i \,\rangle = \epsilon {a_h \over 3} f_i
\label{b7}
\end{equation}
For the zero field limit of the profile we can take the zero field limit
of $f_i$. It has the property that head and tail become equivalent or
\begin{equation} \label{b8}
f_i = 1 - f_{N+1-i} \quad \quad \quad {\rm or} \quad \quad \quad f_i + f_{N+1-i} = 1
\end{equation}
The zero field limit of $f_i$ has been determined in \cite{bark97} by Monte Carlo simulations.
We note that (\ref{b8})  is consistent with the mentioned \cite{next} symmetry
in $\langle y_i \rangle$.  
One should have $a_h=2$ in order that the profile becomes $2 \epsilon /
3$ at the head, as is observed (see Fig. \ref{profE0-001}).  This is
perfectly in agreement with the value $r_h = B^4 \sim 1 + 2 \varepsilon$
for small $\varepsilon$. Combining (\ref{b7}) and the first  Eq. (\ref{bb4})
we find that $f_1 $ is a measure for the drift velocity $v$. According to (\ref{c1}) $f_1$ should
vanish as $1/3N^2$. This result has been derived in  \cite{bark97}. 

Another feature of Fig. \ref{profE0-001} seems to be the collapse of the data
on a single curve. Further data on longer chains show that the flattening--off at the ends
of the chain shrinks with the size of the system and that the slope in the middle slowly decreases.
This is another manifestation of the slow approach towards the asymptotic behavior \cite{next}
for large $N$.

Note that if the division between the head and tail region were located
with equal probability on all sites of the chain then one would have
simply $f_i = i/N$, which from Eq. (\ref{b7}) implies a linear profile.
The profile of Fig. \ref{profE0-001} is linear only at the center of the
chain, while it strongly deviates from linearity close to the edges.
This implies that the probability of finding the division between the
head and tail regions is flat in the center of the chain and drops off
at the chain edges.

\subsection{Long Chains and stronger Fields}

In this case the head zone will be dominant beyond a certain point 
(i.e. $f_i = 1$ for $i > i_0$ ) in
the chain, thus the division between the head and tail zones is expected
to become localized close to the end of the chain which is not pulled.
The curves in Fig. \ref{nk_inter}
convincingly show this behavior. It is interesting to note that when
$f_i \to 1$, Eq. (\ref{b4}) becomes:
\begin{equation} 
\left\{ \begin{array}{rcl}
n^+_i  & = & \displaystyle [1 - n^0_i ] {r_h \over r_h +1} 
\\*[4mm]
n^-_i  & = & \displaystyle [1 - n^0_i  ]{1 \over r_h +1}
\end{array} \right.
\label{l1}
\end{equation}
It immediately implies that both $n^+_i$ and $n^-_i$ are linearly
increasing functions of $i$ in the head zone, being $n^0_i$ a linearly
decreasing function of $i$ (see Eq. (\ref{bb5})). This explains the
monotonic increase of $n^-_i$ close to the pulled end shown in the inset
of Fig. \ref{nk_inter}.  Note that using Eq. (\ref{l1}) one can estimate
$r_h$ from the ratio of the slopes of $n^+_i$ and $n^-_i$ in the head
region. We find a ratio $r_h \simeq B^4$ in agreement with our
crossover formulae.

\section{Discussion}
\label{sec:discussion}

We have presented a series of numerical and analytical results for the
MP problem in the Rubinstein - Duke model, where a single reptating
polymer is pulled by a constant driving field applied to one polymer
end. We have shown that the numerical data for the drift- and curvilinear
velocities can be quite well reproduced by simple interpolating formulas
following from the assumption that both velocities are equal in the limit
of long chains. Indeed the
measured differences are small which shows that the polymer is fairly
stretched by the pulling force.

We studied also local quantities, as the profiles, which provide
information on the shape of the reptating chain. These are quite
well understood using a representation in which the polymer is divided
into a head and tail region, with different ratios of $+$ and $-$ links.
At small fields the division between the two regions meanders trough the 
whole chain, and the probability of finding it close to the edges drops off.
At strong fields the division gets localized close to the free
end of the chain. Moreover some profiles show an unexpected non-monotonic 
behavior which has a simple interpretation in the interface picture.
The precise shape of the profiles at weak fields close to the polymer
edges, both at finite $N$ and in the asymptotic limit $N \to \infty$,
will be discussed in details elsewhere \cite{next}.

{\bf Acknowledgements:}  This work has been supported by the Polish
Science Committee (KBN) under grant in years 2003-2005. A.D. acknowledges 
a grant from the Royal Netherlands Academy of Science (KNAW) enabling him 
to stay at the Leiden University where part of this investigation were 
carried out.

\end{document}